# All-optical polarization control in time-varying low-index films via plasma symmetry breaking

Wallace Jaffray[1], Sven Stengel[1], Alexandra Boltasseva[2], Vladimir M. Shalaev[2], Maria Antonietta Vincenti[3], Domenico de Ceglia[3], Michael Scalora[4], Carlo Rizza[5], Marcello Ferrera[1*]

[1*]Institute of Photonics and Quantum Sciences, Heriot-Watt University, Edinburgh, EH14 4AS, UK.
[2]Elmore Family School of Electrical Computer Engineering and Birck Nanotechnology Center, Purdue University, West Lafayette, Indiana, USA.
[3]Department of Information Engineering, University of Brescia, Brescia, Italy.
[4]Charles M. Bowden Research Center, Redstone Arsenal, Huntsville, Alabama, USA.
[5]Department of Physical and Chemical Sciences, University of L'Aquila, L'Aquila, Italy.

Corresponding authors: m.ferrera@hw.ac.uk;

**Abstract**
Controlling the polarization state of light with sub-picosecond speed and subwavelength precision remains a key challenge for next-generation nano-photonic devices. Conventional methods, such as birefringent crystals, liquid crystals, or electro-optic Pockels cells are limited in speed, compactness and energy consumption. While structured materials and 2D heterostructures show some promise for on-chip ultrafast performance, all-optical control at the nanoscale remains an open issue. Here we introduce an all-optical scheme that uses femtosecond pumping of low-index, subwavelength isotropic films to achieve ultrafast control over birefringence, dichroism, and optical activity in a single platform. When the material is probed at its crossover wavelength, linearly polarized pumping induces a transient phase retardation as large as $0.1\pi$/µm, accompanied by a dichroic absorption ratio of $\approx$ 20 folds. When instead circularly polarized excitation is employed, the probe experiences non-reciprocal optical activity, leading to polarization rotation reaching 1.1°/µm. These transient values are orders of magnitude larger than what is recorded from alternative nanophotonic systems and can be quantitively reproduced by a universal hydrodynamic model attributing these effects to pump-induced symmetry breaking in the photoexcited carrier plasma. Noticeably enough, our model underlines the hidden, but critical role played by the time varying damping parameter towards all polarisation related effects. Our combined experimental and theoretical study establishes a reconfigurable, deep-subwavelength polarization-control mechanism operating on sub-picosecond timescales. This approach is ideally suited for compact ultrafast modulators, dynamic metasurfaces, and tunable nonreciprocal photonic devices, with broad implications for quantum optics, ultrafast logic, and time-resolved sensing.

**Keywords:** Photonics, Nonlinear Optics, Birefringent, Dichroism, Time Varying Optics

# Introduction

Controlling the polarisation state of light with high speed, precision, and reconfigurability is of foundational importance across many fields of modern photonics and optical technologies. Polarisation-sensitive functionalities enables a broad range of applications, from classical and quantum communication to ultrafast optical computing, sensing, and advanced microscopy [1–3]. Integrated polarisation control is particularly vital in planar or on-chip photonic architectures, where conventional bulk elements are unfeasible due to their size, tuning limitations, or speed bottlenecks [4].

Traditional systems such as static birefringent crystals, liquid crystals, and electro-optic Pockels cells enable polarisation transformation but are inherently constrained by slow response times, poor miniaturisation potential, and lack of active programmability [5–7]. To address these limitations, significant effort has been directed at emerging platforms such as structured metasurfaces and van der Waals (2D) heterostructures, to achieve compact and fast polarisation control [8–13]. Metasurfaces, for instance, have shown strong static anisotropy and engineered dichroism through spatial patterning [3, 14–17], while 2D materials and heterostructures offer strong nonlinearities and quantum confinement effects suitable for polarisation-dependent interactions [18, 19]. However, these devices remain limited to a single polarisation mechanism (typically dichroism or birefringence) which is preset by fabrication and often relies upon resonant elements such as plasmonic nanoantennas or dielectric microresonators [20–27]. These resonant dependencies impose constraints on speed and spectral bandwidth, hindering their use in practical ultrafast systems, while the static fabricated structures disable dynamic reconfigurability. Regarding fully-integrated polarisation control, results have been attained from highly nonlinear systems such as colloidal nanoparticles [28] and ultrathin metallic films [29]. Specifically, these systems have been targeting the inverse Faraday effect (IFE) [30–33] to optically induce chiral properties by triggering a static magnetic field and achieving optical activity of the order of $10^{-4} - 10^{-5}$ °/µm.

A new photonic framework has recently emerged from efficient time-varying photonic systems, enabled by ultrafast optical nonlinearities in low-index transparent conductors [34–37]. This marks a fundamental shift in how optical functionalities can be engineered [38]. Rather than relying solely on static, lithographically defined structures to control polarisation (as shown in Fig. 1 upper panels), time has emerged as a powerful new degree of freedom that allows dynamic control over material responses. This transition redefines the paradigm of photonic design, expanding the functional bandwidth and adaptability of optical systems by transferring control from nanofabrication steps towards tailoring the optical excitation signals. Within this context, ultrafast polarisation control has been theoretically discussed in several works [39–42].

Figure 1 provides a conceptual comparison between conventional metasurface-based approach and temporally modulated materials. In the structural solution, each functionality (e.g., birefringence, dichroism, chirality) requires a distinct geometrical design, while also being reciprocal (see input/output signal icon below each device). In contrast, the temporal approach allows a single unstructured material platform to achieve multiple optical tasks with dynamic tunability, by simply shaping the excitation beam. This enormous technological adaptability comes together with non-reciprocal behaviour which can be used for key operations such as optical isolation or non-reciprocal phase shifting [43].

Here, we present a fundamentally different approach for universal, ultrafast, and reprogrammable polarisation control. In our platform, birefringence, dichroism, and chirality are optically induced and imprinted onto a nonlinear low-index medium via shaping a femtosecond optical pump. Our system is based on subwavelength transparent conducting oxide (TCO) film operated near their epsilon-near-zero (ENZ) point. Using linearly and circularly polarised pump beams at variable delay from a weak probe, we demonstrate all-optical activation and temporal control over the three fundamental polarisation interactions, namely, birefringence, dichroism, and optical activity on sub-picosecond timescales. Under linearly polarised pumping, we report induced birefringence $\Delta\phi_L \approx 0.1\pi$/µm (here defined as nonlinear phase-retardation per unit length between orthogonal polarisation components) and a dichroic absorption ratio $P_x/P_y \approx 20\%$ (where $P_x$ and $P_y$ are the measured probe output powers along the orthogonal reference axis). Subsequently, under circularly polarised excitation, we observed induced optical activity manifested in the polarisation rotation of the transmitted linearly polarised probe in both clockwise and counterclockwise directions, depending on the pump helicity, with rotation rates reaching $\pm 1.1$°/µm. These results experimentally realise key concepts which were previously proposed in theoretical frameworks, such as dynamic chiral modulation in time-varying permittivity media [44, 45].

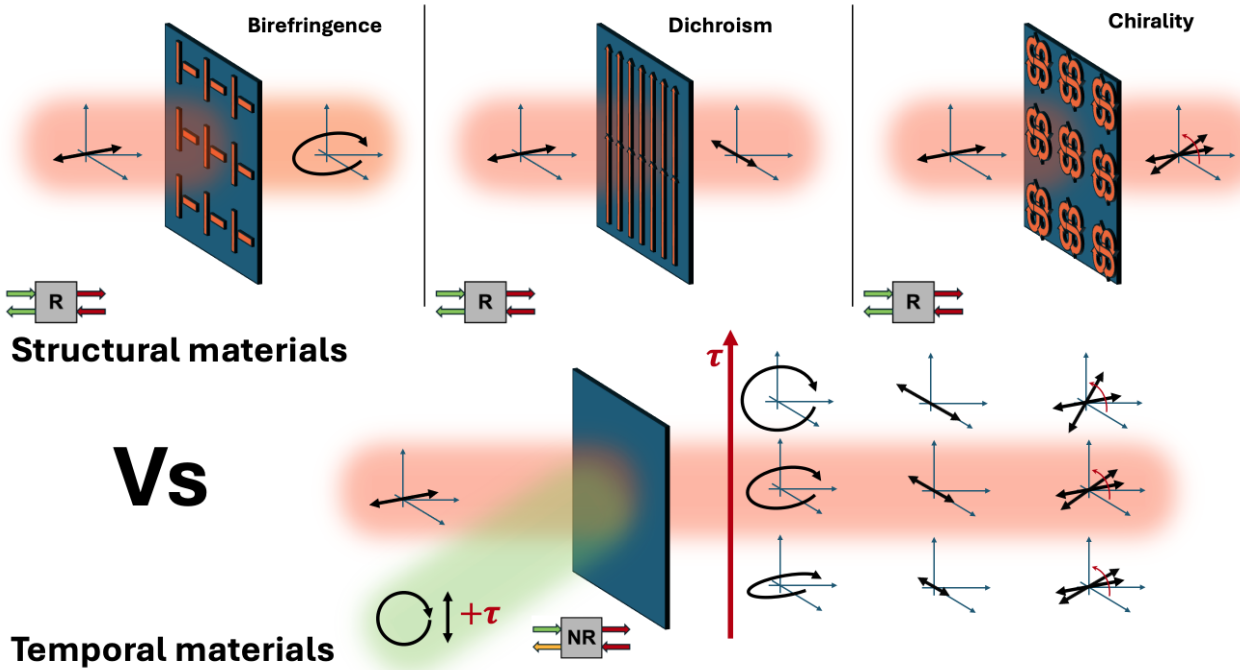

**Fig. 1 Conceptual comparison between structural and temporal approaches to polarisation control.** (Top) Structural metasurfaces employ geometrically distinct designs to achieve reciprocal (R) birefringence, dichroism, or chirality, each requiring separate fabrication. (Bottom) In contrast, temporal materials respond to tailored optical excitations, where the time profile and polarisation of the incident beam dynamically induce birefringent, dichroic, or chiral behaviours in a single unpatterned substrate. This temporal engineering approach greatly enhances reconfigurability, reduces fabrication overhead, and expands the accessible optical response space while also enabling non-reciprocal (NR) functionalities.

To interpret these phenomena, we develop an effective model that captures the temporal evolution of the material's anisotropic electron distribution and resulting dielectric tensor asymmetry. This model reproduces all observed polarimetric responses with good accuracy and links the symmetry-breaking mechanism to ultrafast non-equilibrium carrier dynamics, a feature uniquely enabled by the absence of geometric anisotropy or fabrication-imposed symmetry.

Critically, our system achieves polarisation reconfigurability without relying on resonance, which enables broadband operation and extreme temporal agility. To the best of our knowledge, this is the first demonstration of simultaneous access to all three major polarisation control mechanisms on a single material platform, achieved purely through temporal engineering of the pump field. This decoupling of spatial fabrication and temporal functionality introduces a new paradigm for reconfigurable nanophotonics and opens pathways to ultrafast polarisation modulators, dynamic metasurfaces, optically addressable nonreciprocal devices, and ultrafast computing in both classic and quantum regimes [46–48].

## Experimental settings

Experiments were conducted using a pump–probe setup designed to investigate the ultrafast modulation of polarisation states in a time-varying, subwavelength film of TCO. The sample used was a 900 nm thick aluminium zinc oxide (AZO) film deposited via pulsed laser deposition onto a fused silica substrate in a low-oxygen environment [38, 49]. The optical response of the unpumped film is near-isotropic and exhibits negligible static birefringence or dichroism, making it ideal for isolating nonlinear time varying polarisation effects. The probe beam was supplied by an optical parametric amplifier (OPA) set to an output wavelength of 1250 nm, corresponding to a wavelength slightly below the ENZ crossover point of the AZO film. The pump beam was fixed at 787 nm with a pulse duration of 100 fs, a repetition rate of 10 Hz, and a peak intensity of approximately 800 GW cm$^{-2}$. The beams were aligned in a near-collinear geometry and incident on the sample at close to normal incidence ($< 5°$). The pump was polarised linearly along $0°$ (horizontal and aligned with the plane of incidence), with a beam diameter of ∼1.5 mm ($1/e^2$), considerably larger than the probe beam diameter (∼200 µm), ensuring uniform excitation of the probed region. The probe intensity was set several orders of magnitude below the pump intensity to avoid self-action nonlinearities.

Polarisation-resolved measurements of the transmitted probe were obtained by placing a motorized analyser along the transmitted beam path, after a frequency filter dedicated to remove any residual pump, and then measuring the output power at discrete angular steps of $2°$. All data were averaged over multiple laser shots to improve the signal-to-noise ratio. For a first set of measurements pertaining optically induced birefringence and dichroism, we used three input probe polarisation states: $0°$, $45°$, and $90°$ (i.e., horizontal, diagonal, and vertical with respect to the horizontal pump polarisation), while varying the pump-to-probe delay $\tau$ in steps of about 10 fs (see Fig. 2). Subsequently, to investigate the induced optical chirality, the probe was set to a vertical polarisation state (90°) while the pump was alternatively arranged into right circularly polarised (RCP) and left circularly polarised (LCP) states. All these configurations are reported at the bottom of Fig. 2, where input polarisation status for the pump and probe are shown in the lower panels. For the rest of this manuscript we assume for all polarisation diagrams that light propagates inwards towards the display plane.

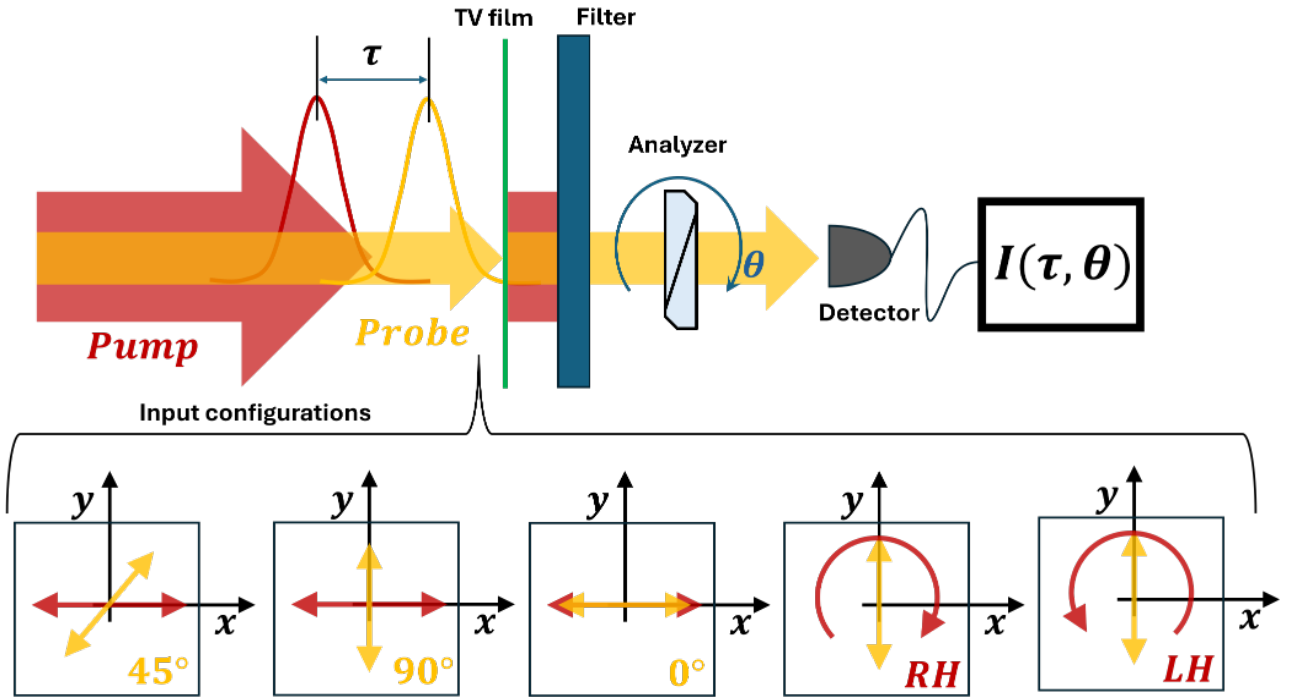

**Fig. 2 Experimental settings.** A time-varying subwavelength thin film is attained by optically pumping a 900 nm of aluminium zinc oxide (AZO) layer via a 10 Hz train of 100 fs pulses centered at 1250 nm, which is within the ENZ bandwidth of the material. Panels along the bottom display the various input experimental configurations of pump and probe mutual polarisations. The output intensity is measured as a function of pump-probe delay and analyser angle.

The observed polarisation transformations were analysed in terms of induced birefringence (phase retardation under linear pumping), dichroism (differential absorption under linear pumping), and optical activity (polarisation rotation under circular pumping), quantified from the angle and time dependant probe intensity $I_{exp}(\tau,\theta)$ as recorded at the detector located in front of the analyser.

## Results and discussion

### Transient birefringence and dichroism

With the probe polarisation at 45° and the pump polarisation at 0° (parallel to the plane of incidence), we can recover both the induced transient birefringence and dichroism. Figure 3a provides the raw experimental data for this case, where $I_{\text{exp}}(\tau,\theta)$ is the recorded probe intensity as a function of pump-probe delay $\tau$ and analyser angle $\theta$. At early delays, in the absence of pumping, the probe maintains its linearly polarised state after transmission, and the analyser scan leads to a symmetric cosine-squared profile with no angular shift, as expected from Malus's Law. At subsequent delays, where the probe is overlapped with the pump, there is an increase in transmission, and a shift in the center of mass of $I_{\text{exp}}(\tau,\theta)$. To further analyse this process we can mathematically represent the measured intensity after both the time varying film and the analyser as follows:

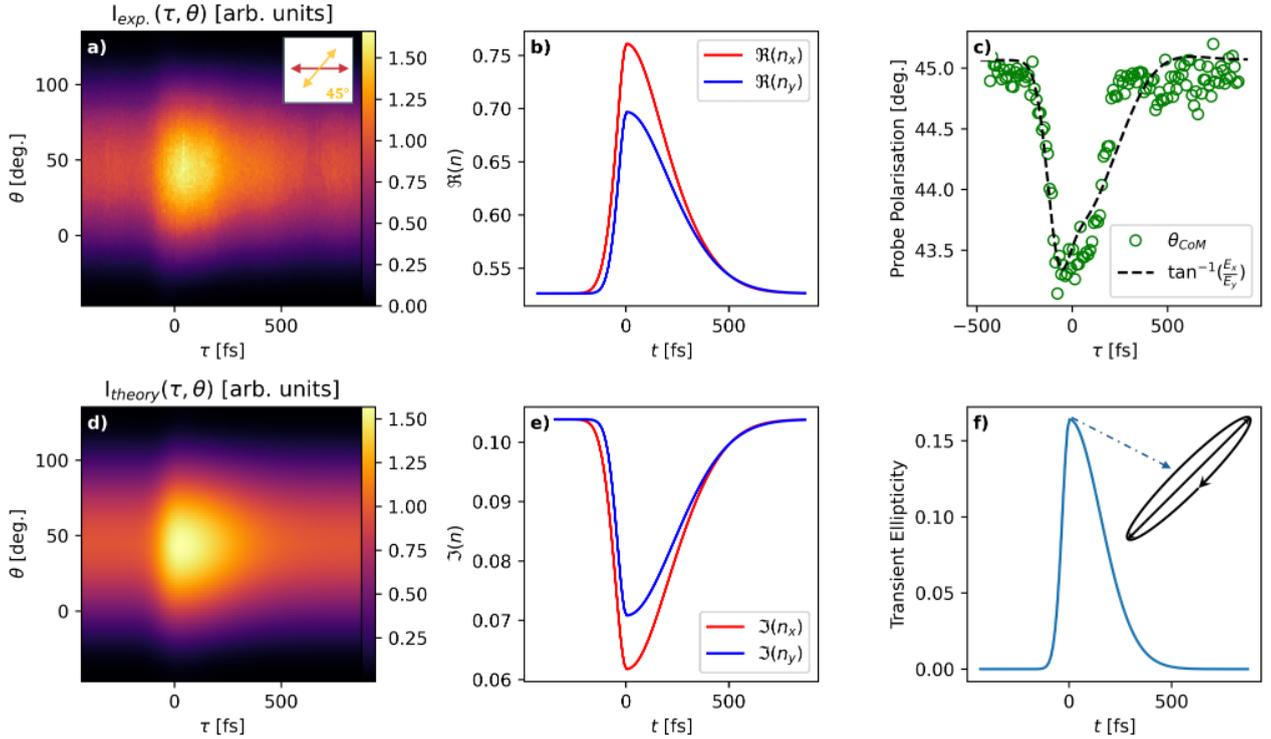

**Fig. 3 Ultrafast nonlinear polarisation coupling in a time-varying AZO film.** Experimentally measured and theoretically reconstructed polarisation-resolved probe transmission through time-varying, 900-nm-thick aluminium zinc oxide (AZO) layer. **a)** Raw experimental intensity $I_{\text{exp}}(\tau,\theta)$ as a function of analyser angle $\theta$ and pump–probe delay $\tau$, with the probe polarised at 45° and the pump at 0°. **b)** Real part of the transient refractive index retrieved from fitting the measured intensity, showing anisotropic changes along the $x$ (red) and $y$ (blue) directions. **c)** Extracted polarisation angle of the transmitted probe from both experimental center-of-mass analysis (green circles) and model (black dashed line), demonstrating excellent agreement. The theoretical evaluation is attained by using the formula reported in the inset, where the angled brackets indicate a time average integral over absolute time $t$. **d)** Simulated intensity map based on fitted complex indices, accurately reproducing the measured transmission with RMS error below a few percent. **e)** Imaginary part of the refractive index as retrieved from fitting, showing transient, polarisation-dependent absorption. **f)** Retrieved ellipticity of the transmitted pulse indicating the onset of nonlinear birefringence-induced elliptical polarisation; inset shows the reconstructed polarisation ellipse at maximum anisotropy, with ellipticity approaching a 1:7 ratio.

$$I_{\text{exp}}(\tau,\theta) = \int_{-\infty}^{\infty} [\cos^2(\theta)|E_x(t,\tau)|^2 + \sin^2(\theta)|E_y(t,\tau)|^2$$
$$+ 2|E_x(t,\tau)||E_y(t,\tau)|\sin(\theta)\cos(\theta)\cos(\delta(t,\tau))]dt \qquad (1)$$

Where $E_x(t,\tau)$ and $E_y(t,\tau)$, are the instantaneous, transmitted electric fields in the x-direction (0°) and y-direction (90°), respectively, as a function of absolute time $t$ and at a given pump-probe delay $\tau$. Here, $\delta(t,\tau)$ is the phase difference between $E_x$ and $E_y$, which relates to the transient birefringence of the material. The integral over time deals with the power acquisition over the pulse duration. As the temporal profile of our input probe pulse is known (evaluated via FROG measurements), we can fit Eq. 1 to the intensity heatmap in Fig. 3a to recover a transient complex index in the $x$ and $y$ directions. Figure 3d shows the fitted heatmap, which closely matches the experimentally acquired data with less than a few percent RMS error. The real and imaginary components of the time varying index are provided in Fig. 3b and e, respectively. Here, the red lines (both solid and dashed) show the complex index as perceived by the horizontal field components of the probe, while the blue lines show the complex index in the y-direction. The recovered temporal index is remarkably aligned with previous measurements of time varying index in AZO [50].

To further verify the accuracy of our experimental fit, and our understanding of the nonlinear process, we calculate the polarisation angle of the transmitted light as a function of the pump-probe delay. This is done using the previously evaluated complex index (Fig. 3b and e) and a temporal average over the absolute time $t$ (shown as the black dotted line in Fig. 3c). This is then compared to the center of mass of the experimental data shown as the green circles in Fig. 3c. This provides us with a direct estimate of the polarisation angle of the transmitted light, which matches closely with the theoretical treatment (see material model section). Additionally, it is worth noting that the polarisation of the transmitted light is rotated towards the horizontal pump by 1.5° after propagating only 900 nm. We can also recover the transient birefringence of the film, which imparts ellipticity onto the transmitted pulses as shown in Fig. 3f. Specifically, as the probe interacts with the time varying film, it gains a substantial ellipticity up to a ratio of almost 1:7 due to the induced difference between orthogonal index components (see to-scale ellipse in inset of Fig. 3f).

Figure 4 illustrates the case of a vertically polarised probe (top row) and horizontally polarized probe (bottom row), both for a horizontally polarized pump excitation. As in previous examples, panels 4a and d show the experimentally measured intensity maps, while panels 4b and e provide the corresponding theoretical fits using

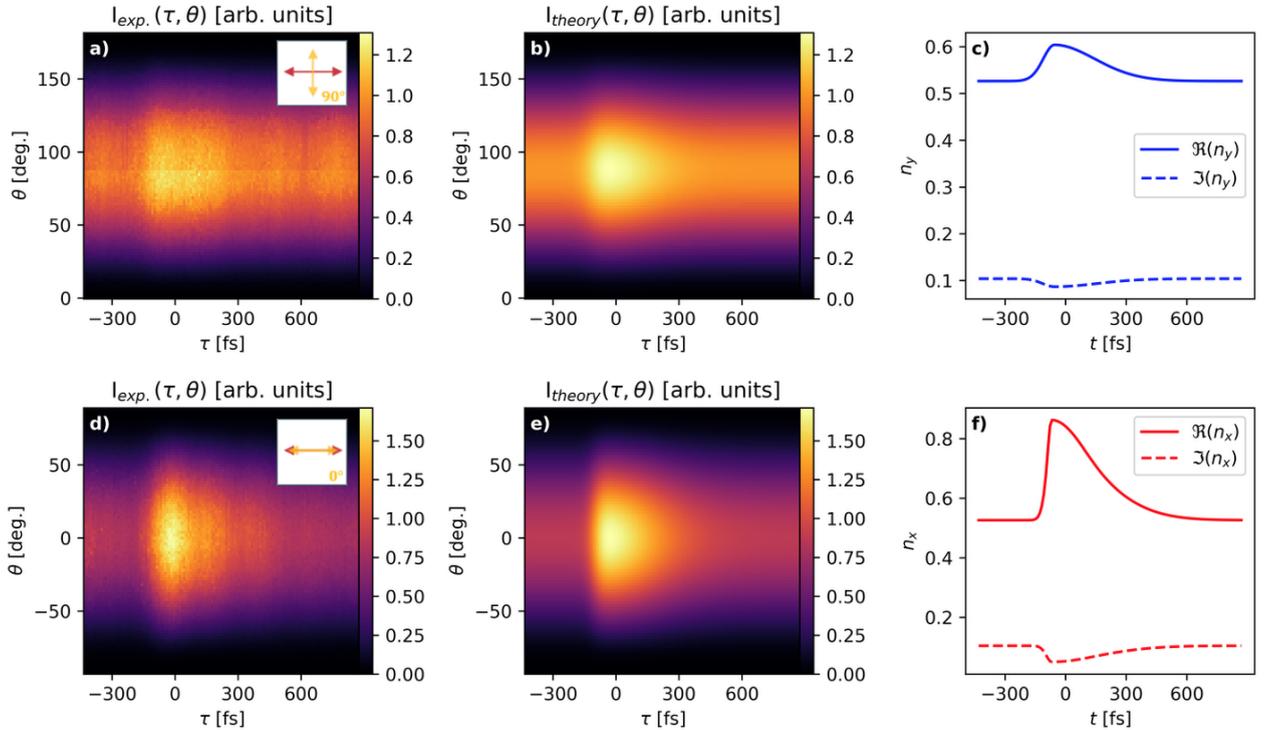

**Fig. 4 Polarisation-resolved probe transmission under a fixed horizontal pump with either vertical (top row) or horizontal (bottom row) probe polarisation. a)** Experimentally measured intensity map for a vertically polarised probe. **b)** Fitted intensity distribution for vertically polarised probe. **c)** Extracted transient complex index in the $y$ direction (orthogonal to the pump). **d)** Experimentally measured intensity map for a horizontally polarised probe. **e)** Extracted transient complex index in the $x$ direction (in line with pump). **c)** Extracted transient complex indices in the $x$ direction (along the direction of the pump)

our model. In this measurement configuration, the model accurately retrieves only the complex refractive index along the specific direction of the probe polarisation. The extracted complex indices are shown in panels 4c and f. These measurements reveal key aspects of the nonlinear coupling between polarisation components. When the time-varying refractive index is probed independently along the x- and y-directions, we observe a stronger nonlinear response in the x-direction and a weaker response in the y-direction when compared to the isolated components from the 45° polarised probe case. This asymmetry cannot be explained by a simple vectorial decomposition of the linear responses, indicating the presence of nonlinear inter-component energy transfer.

While this nonlinear coupling is explained in more details by our material model, a more intuitive physical interpretation can be helpful. In this regard, the dominant nonlinear mechanism in the studied AZO films is the hot electron effect, which modifies the effective mass of the conduction electrons (effectively forming an ultrafast light-driven plasma) [51]. Under optical excitation the pump increases the electron momentum along its polarisation direction, driving electrons up into the conduction band. These hot carriers subsequently scatter, transferring energy to the lattice (electron–phonon scattering) and to other electrons (electron–electron scattering), with the material quickly returning to its linear optical state. Crucially, electron–electron scattering redistributes the momentum imparted by the pump to off-axis directions. This alters the off-axis effective carrier mass, and consequently the complex refractive index in those directions. The angular dependence of this scattering process, governed by a Yukawa-screened Coulomb potential, exhibits a peak scattering amplitude between 20° and 50° relative to the initial pump polarisation direction [52, 53]. This theoretical prediction is consistent with our experimental observation where the change $n_y$ under 45° probe polarisation is stronger than measured with a vertically polarised probe.

## Induced chirality

To investigate induced optical activity, we replace the linearly polarised pump with circularly polarised excitation, while keeping the probe fixed at a vertically polarised state (90°). By switching between right-handed (RCP) and left-handed (LCP) circular polarisation of the pump, we examine the induced transient chirality and its non-reciprocal properties in an otherwise achiral and isotropic system.

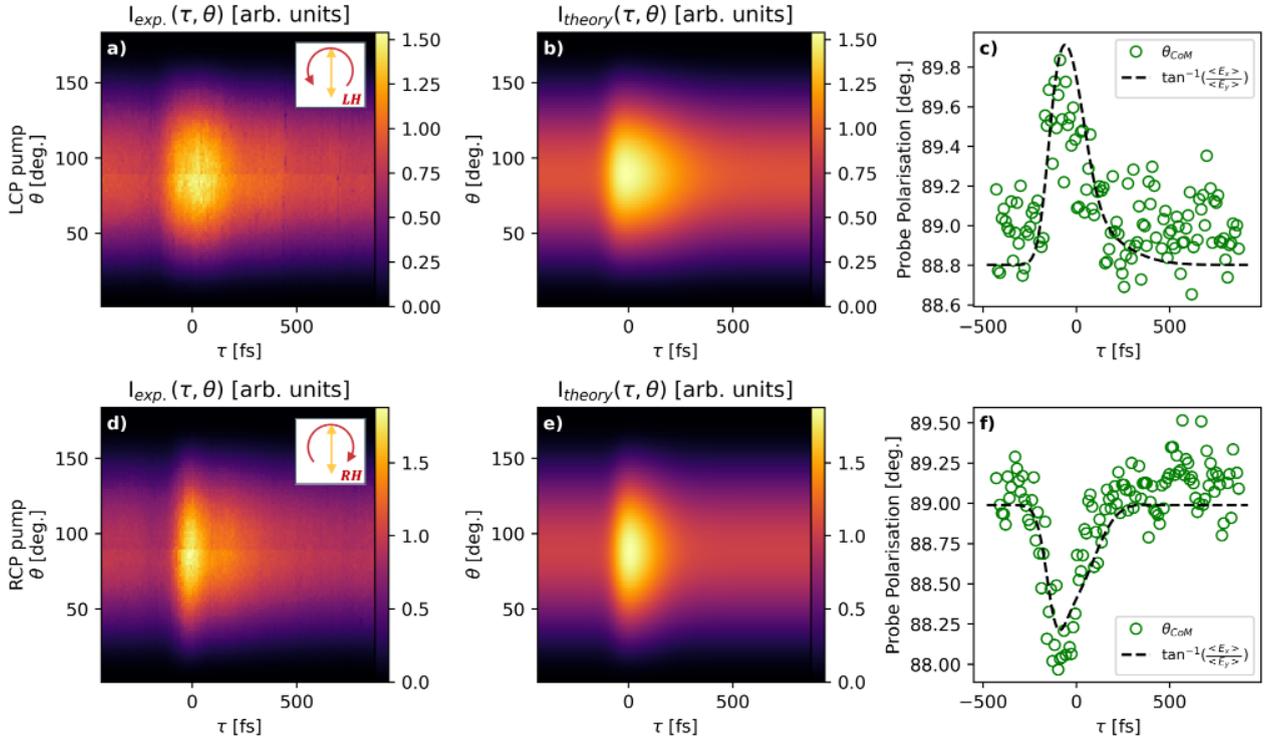

**Fig. 5 Transient polarisation rotation of a vertically polarised probe induced by circularly polarised pump pulses in an isotropic AZO film. a)** Measured probe transmission as a function of analyser angle $\theta$ and pump–probe delay $\tau$ for a left-handed circularly polarised (LCP) pump. **b)** Theoretical fit to the probe's measured intensity distribution for a LCP pump. **c)** Polarisation rotation angle for an LCP pump, which is extracted from the analyser-scan Centre of Mass ($\theta_{CoM}(\tau)$ and green circles) and from the theoretical fit (black dashed line), showing excellent agreement. The theoretical evaluation is attained by using the formula reported in the inset, where the angled brackets indicate a time average integral over absolute time $t$. The rotation reaches peak values of $\pm 1°$, corresponding to a nonlinear rotation rate of approximately 1.1°/µm. **d)** Measured probe transmission for a right-handed circularly polarised (RCP) pump. **e)** Theoretical fit to the probe's measured intensity distribution for a RCP pump. **f)** Polarisation rotation angle $\theta_{CoM}(\tau)$ for a RCP pump, which is extracted from the analyser-scan centre of mass (green circles) and from the theoretical fit (black dashed line). As expected, the RCP pumped nonlinearity has the opposite effect to the LCP case.

Figure 5a and d show the transmitted probe intensity as a function of analyser angle and pump–probe delay for a LCP pump (top row) and RCP pump (bottom row). As above, in the absence of pumping the polarisation of the transmitted probe is unchanged. However, upon optical excitation with circularly polarised pump pulses, the probe emerges with a rotated polarisation axis whose direction depends on the pump helicity. This asymmetric optical activity is signature of optically induced chirality.

To quantify the effect, we complete the same analysis as in the previous section, with fitted intensity profiles $I_{theory}(\theta,\tau)$ now shown in Fig. 5b and e. The instantaneous polarisation angle is provided in Fig. 5c and f, and can be directly extracted from both $I_{\exp}(\theta,\tau)$ (green circles) and from our fit (dashed black lines). Our theoretical evaluation is calculated using $\tan^{-1}\left(\frac{\langle E_x\rangle}{\langle E_y\rangle}\right)$, where the angled brackets indicate a time average integral over the absolute time $t$. The polarisation axis rotates clockwise for RCP and counterclockwise for LCP, reaching peak values of approximately $\pm 1°$ within the duration of the pump–probe overlap. This corresponds to a non-reciprocal rotation rate of $\approx 1.1°/\mu m$, which is several orders of magnitudes higher than those achieved in other integrated systems [28, 29].

Unlike natural optical activity, which arises from intrinsic material chirality and is fixed in sign and magnitude, the chiral response here is dynamically induced and tunable. The mechanism stems from an asymmetric redistribution of carriers in the presence of circularly polarised electric fields, which transiently breaks mirror symmetry in the electron plasma. This leads to off-diagonal components in the time-dependent permittivity tensor, and thus to nonreciprocal coupling between orthogonal polarisation components of the probe.

## Material model

In our material model, the intrinsic time variation of the effective mass alters the plasma frequency and scattering rate [54]. Despite being often overlooked in the literature, the latter plays a central role in nonlinear polarisation coupling. Specifically for the case of a circularly polarised pump, our treatment unveils the hidden, but fundamental, link between the transient behaviour of the damping parameter with the emergence of a remarkable IFE [30–33]. Returning to our specific experimental configuration, where the probe intensity is much lower than that of the pump, the dynamics of the probe can be represented by the following oscillator equation modified for a time varying damping and plasma frequency:

$$\frac{\partial^2 \boldsymbol{\mathcal{P}}_2}{\partial t^2} + \overset{\leftrightarrow}{\gamma}_2(z,t)\frac{\partial \boldsymbol{\mathcal{P}}_2}{\partial t} = \varepsilon_0 \omega_p^2 \overset{\leftrightarrow}{F}_2(z,t)\boldsymbol{\mathcal{E}}_2, \qquad (2)$$

where $\boldsymbol{\mathcal{P}}_2(\mathbf{r},t)$ is the polarization vector driven by the probe field, $\boldsymbol{\mathcal{E}}_2(\mathbf{r},t)$ is the electric field of the probe, $\omega_p$ is the plasma frequency, $\epsilon_0$ is the vacuum permittivity, $\overset{\leftrightarrow}{\gamma}_2(z,t)$ is a nonlinearly modified time varying scattering rate, and $\overset{\leftrightarrow}{F}_2(z,t)$ is a time dependant modification to the plasma frequency. These latter two values are time varying bianisotropic complete tensors, represented by the overset arrow.

Figure 6 presents the results of full-wave simulations with the coupled equation described in Eq. 2 for both previous studied cases, where the material was linearly and circularly pumped. In the first scenario the pump wavepacket is linearly polarised along the x-axis while the probe wavepacket is linearly polarised at $45°$, leading to an induced temporal anisotropy and birefringence. Panel a) of Fig. 6 displays the spectral components of the transmitted electric fields at the probe frequency, $\omega_2$. More specifically, the electric fields of the probe in the x- and y-directions ($|\tilde{E}_{t,x}(\omega_2)|$ and $|\tilde{E}_{t,y}(\omega_2)|$) are plotted as a function of the delay time $\tau$. Fig. 6 b) illustrates the polarization rotation angle, given by $\tilde{\theta}_t(\omega_2) = \tan^{-1}\left[\frac{|\tilde{E}_{t,y}(\omega_2)|}{|\tilde{E}_{t,x}(\omega_2)|}\right]$. In the second configuration, the pump beam is

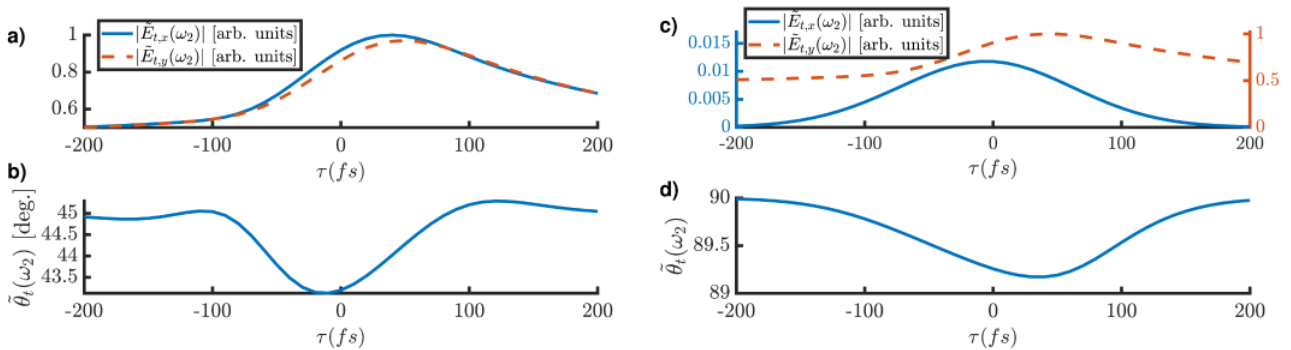

**Fig. 6 Full-wave simulations with coupled equations.** The spectral $x$, and $y$ components of the transmitted electric fields at frequency $\omega_2$ are shown in panel (a) as functions of the delay time $\tau$, while panel (b) displays the polarization rotation angle $\tilde{\theta}_t(\omega_2)$. This is for the case where the pump wave packet is linearly polarized along the $x$-axis (i.e., $\theta_1 = 0°$ and $\delta_1 = 0°$), and the probe wave packet is linearly polarized with $\theta_1 = 45°$ and $\delta_1 = 0°$. (c,d) Similar to panels (a,b) but with the pump beam being circularly polarized with $\theta_1 = 45°$ and $\delta_1 = 90°$, while the probe beam being linearly polarized along the $y$-axis (i.e., $\theta_2 = 90°$ and $\delta_2 = 0°$).

right handed circularly polarised, while the probe beam is linearly polarized along the y-axis, thus inducing a transient chirality. This is shown in Fig. 6c and d, where the probe electric field and output polarisation angle are shown as a function of time delay $\tau$. Theoretical results for both linear and circular pump polarisations show good agreement with the experimental results.

# Conclusion

We have demonstrated an all-optical tunable, ultrafast, and reconfigurable platform for controlling the polarisation state of light by dynamically inducing birefringence, dichroism, and non-reciprocal chirality in an unstructured, planar, and subwavelength system. Using femtosecond laser excitation of an AZO film within its ENZ wavelength regime, we optically induce transient anisotropies in the material's dielectric response that enable all the three key polarisation interaction mechanisms. Notably, this polarisation control is achieved without any reliance on resonant structures or static fabrication-induced symmetry, allowing for broadband operation and extreme temporal agility. Linearly polarised pumping generates transient birefringence and anisotropic absorption, with observed phase shifts of up to $0.1\pi/\mu m$ and dichroic ratios reaching 20%. Circularly polarised pumping, in contrast, introduces dynamically reconfigurable optical activity, with polarisation rotation exceeding $1°/\mu m$, which is a clear manifestation of light-induced chirality in an otherwise isotropic medium. These phenomena were quantitatively captured by a first-principles model that attributes the observed effects to ultrafast, symmetry-breaking electron dynamics in the photoexcited plasma.

This work not only establishes a new paradigm for ultrafast polarisation modulation via time-varying media but also sets the foundations for a class of compact, optically programmable photonic devices. The demonstrated compatibility with planar integration and femtosecond response opens up new avenues for applications in ultrafast optical switching, quantum information processing, dynamic metasurfaces, and nonreciprocal photonic systems